\documentclass[a4paper,11pt]{article}
\pdfoutput=1
\usepackage{graphicx}
\usepackage{epsf,amsmath,bbold,amsfonts,stmaryrd}

\usepackage[utf8]{inputenc}
\usepackage{mathrsfs}
\usepackage{appendix}
\usepackage{amssymb}
\usepackage{float}
\usepackage{color}
\usepackage{cite}
\usepackage{hyperref}
\hypersetup{pageanchor=false}
\usepackage{indentfirst}
\usepackage{url}
\usepackage{float}
\usepackage{caption}
\usepackage[numbers,square,comma,sort&compress,merge]{natbib}

\def\mc{\mathcal}

\def\nn{\nonumber}

\def\be{\begin{equation}}
\def\ee{\end{equation}}

\def\bea{\begin{eqnarray}}
\def\eea{\end{eqnarray}}

\def\ba{\begin{array}}
\def\ea{\end{array}}

\def\bc{\begin{center}}
\def\ec{\end{center}}

\def\bl{\begin{flushleft}}
\def\el{\end{flushleft}}

\def\br{\begin{flushright}}
\def\er{\end{flushright}}

\def\bi{\begin{itemize}}
\def\ei{\end{itemize}}

\def\bt{\begin{tabular}}
\def\et{\end{tabular}}
\newcommand{\sR}{\mathsf{R}}
\newcommand{\mR}{\mathcal{R}}

\numberwithin{equation}{section}

\begin{document}
\title{\textbf{Photon Emission Near Myers-Perry Black Holes in the Large Dimension Limit}}
\author{Minyong Guo$^{1}$\footnote{minyongguo@pku.edu.cn}~,
Peng-Cheng Li$^{1,2}$\footnote{lipch2019@pku.edu.cn, corresponding author}~
and
Bin Chen$^{1,2,3}$\footnote{bchen01@pku.edu.cn}}
\date{}

\maketitle

\vspace{-10mm}

\begin{center}
{\it
$^1$Center for High Energy Physics, Peking University,
No.5 Yiheyuan Rd, Beijing 100871, P. R. China\\\vspace{1mm}

$^2$Department of Physics and State Key Laboratory of Nuclear
Physics and Technology, Peking University, No.5 Yiheyuan Rd, Beijing
100871, P.R. China\\\vspace{1mm}

$^3$ Collaborative Innovation Center of Quantum Matter,
No.5 Yiheyuan Rd, Beijing 100871, P. R. China
}
\end{center}

\vspace{8mm}

\begin{abstract}
We study the null geodesics extending from the near-horizon region out to the far region in the background of the
Schwarzschild and the singly-spinning Myers-Perry black holes in the large dimension limit. We find that in this limit the radial integrals of these geodesics can be obtained by using the method of matched asymptotic expansions. If the motion of the photon is confined to the equator plane, then all geodesic equations are solvable analytically. The study in this paper may provide a toy model to analyze the observables relevant to the electromagnetic phenomena occurring near the black holes.
\end{abstract}

\maketitle

\newpage

\section{Introduction}
The first image of the supermassive black hole at the centre of the galaxy M87, was observed by the Event Horizon Telescope (EHT) recently \cite{Akiyama:2019cqa, Akiyama:2019brx,Akiyama:2019sww,Akiyama:2019bqs,Akiyama:2019fyp,Akiyama:2019eap}, which opens another new window to the study of black holes apart from gravitational wave detectors such as LIGO and Virgo. Owing to the development of new techniques in the future, the EHT is expected to give more accurate data and thus takes our understanding of black holes to a higher level experimentally. At the same time, the theoretical research  needs  urgently to be further developed  in many aspects as well, especially on the investigation of the null geodesics and the shadow of a black hole.

As it is well known, the equations of null geodesics from the near horizon to the infinity cannot be solved analytically in general, which makes it difficult to obtain analytically the time and the rotational angle of the photon along the geodesic trajectory. Remarkably, in an instructive paper \cite{Porfyriadis:2016gwb}, the authors found that in the background of the extremal Kerr black holes the radial integrals of the geodesic equations can be performed analytically to the leading order in the deviation from the superradiant bound. This is achieved by using the method of matched asymptotic expansions (MAE), since the emergence of an enhanced conformal symmetry in this case can greatly simplify the gravitational dynamics in the near-horizon region. It was also suggested that by solving the null geodesics for an observer far from the black hole using the the geodesic optics methods of \cite{Cunningham:1973, Cunningham:1975}, one can get analytical formulas and may have a deeper understanding about observational properties of electromagnetic radiation. Furthermore in \cite{Gralla:2017ufe}, by applying the analytical results of null geodesic equations, the authors  predicted  the observational  signature of the high spin black holes at the EHT such as the image positions and the fluxes.\footnote{For the related works, one can see \cite{Lupsasca:2017exc,Wei:2016avv, Guo:2018kis, Gates:2018hub, Long:2018tij, Yan:2019etp,Igata:2019pgb,Kapec:2019hro,Igata:2019hkz,Guo:201911}.} However, these studies are based on the enhanced symmetry in the near-horizon geometry which exists only for the (near-) extremal spinning black holes, thus cannot apply to the rotating black hole of a general spin.

In this paper, we would like to study the null geodesics in the background of the rotating black holes  in the large dimension limit.  When the dimension of the spacetime becomes large enough the dynamics of black holes can be dramatically simplified \cite{Emparan:2013moa}. Physically, this simplification can be attributed to the fact that the gravitational field is confined to the near-horizon region  of the black hole. Similar to the extremal Kerr black holes,  in this limit the black holes possess well-defined near-horizon regions with universal features and enhanced symmetry \cite{Emparan:2013xia}. Consequently, the large $D$ expansion method can be used to calculate the quasinormal modes of  the gravitational perturbations of the black holes analytically \cite{Emparan:2014cia, Emparan:2014jca, Emparan:2014aba,Chen:2015fuf}. Moreover, the studies suggest that the perturbation about a black hole could be classified into two kinds in the large $D$ limit: the heavy ones which is universal and featureless, and the light one which captures the instabilities of the black hole \cite{Emparan:2014jca}. The class of light modes decouples from the asymptotic region such that one can formulate  an effective theory of  black hole \cite{Emparan:2015hwa,Bhattacharyya:2015dva}. The effective theory  has been widely applied to study of various  issues on black hole, and has also been developed for the black holes in the Gauss-Bonnet gravity \cite{Chen:2017hwm}. 

The aim of this paper is to obtain the analytical results for the radial integrals of the null geodesic equations to the leading order in the inverse of the large spacetime dimension (i.e. $1/D$) by the means of MAE. Taking the large $D$ limit, the intense gravitational field plays a dominant role in the near-horizon region while in the far region the spacetime can be treated as a flat spacetime which makes the calculations simplified in both regions separately. Moreover, we find the integration constant of each side matches very well in the overlap region (the definitions of different regions will be given in the following). As a result, we can get algebraic equations which relate the starting point of the null geodesic near the horizon to the end point in the region far from the black holes if we constrain the null particles to the equatorial plane. It is worth mentioning that for the Myers-Perry black holes, the stable condition for the gravitational perturbations must be taken into account. In contrast,  the Schwarzschild black hole is always stable \cite{Suzuki:2015iha}.

From the practical point of view, we have to admit that the present work is less-than-desirable when the dimension of the spacetime is not large enough. The shortcoming might be remedied by pushing the computation to higher orders in $1/D$. On the other hand, compared to the results in \cite{Porfyriadis:2016gwb}, the main advantage of our treatment is that the spin of the black holes  is general. Thus when we study the observational problems which involve the electromagnetic radiation  emitting in the near-horizon region, it will be very useful to investigate the universal properties of the influences of the spin. It might be possible that some properties are valid even for four-dimensional rotational black hole such that the work may provide a  model to match the experimental data from the EHT.

We organize this paper as follows. As a warm up, in section \ref{Sch_BH} we introduce our framework in details and perform the radial integration to the leading order in $1/D$ for the Schwarzschild black holes.  In section \ref{MP_BH}, we move to the singly-spinning Myers-Perry black holes. We give a summary in section \ref{summary}. In appendix \ref{appA}, we present some specific calculations related to Eqs. (\ref{I1}) and (\ref{I2}).

\section{Schwarzschild black hole}\label{Sch_BH}

\subsection{Geodesic motion}
The metric of a $D$ dimensional Schwarzschild black hole is given by
\be
ds^2=-\left(1-\frac{r_0^{D-3}}{r^{D-3}}\right)dt^2+\frac{dr^2}{1-\frac{r_0^{D-3}}{r^{D-3}}}+r^2(d\theta^2+\sin^2\theta d\phi^2+\cos^2\theta d\Omega^2_{D-4}),
\ee
where $r_0$ is the horizon radius of the black hole and $d\Omega^2_{D-4}$ is the line element of a unit $(D-4)$ sphere.

The motion of a particle is governed by the Hamilton–Jacobi equation
\bea
\frac{\partial S}{\partial \sigma}+\frac{1}{2}g^{ab}\partial_aS\partial_bS=0,
\eea
and the conjugate momentum is defined by $p_a=\partial_aS$, thus the Hamiltonian for the geodesic motion in the spacetime is given by
\be
H=\frac{1}{2}g^{ab}p_a p_b=-\frac{1}{2}m^2.
\ee
Obviously, for null particles, we have $m=0$ and $H=0$ therefore. In addition, there are other two conserved quantities for such null geodesics, i.e.
\bea\label{ce}
E&=&-p\cdot \partial_t=-p_t,\\\label{cl}
L&=&p\cdot  \partial_\phi=p_\phi,
\eea
which represent the energy and angular momentum of the photon.

Plugging Eq. (\ref{ce}) and (\ref{cl}) into the expression of the Hamiltonian $H$ and assuming that $p_\theta$  depends only on $\theta$ and $p_r$  depends only on $r$, we find that the Hamiltonian gives the relation
\be
p_\theta^2+\frac{L^2}{\sin^2\theta}=
-\left(1-\frac{r_0^{D-3}}{r^{D-3}}\right)r^2 p_r^2+\frac{r^2E^2}{1-\frac{r_0^{D-3}}{r^{D-3}}}-r^2m^2=\kappa,
\ee
where $\kappa$ is a constant and is related to a conserved quantity, i.e. the Carter constant \cite{Carter:1968rr} by
\be
Q= \kappa-L^2=p_\theta^2+p_\phi^2\cot^2\theta.\label{cq}
\ee
By now there are four conserved quantities for the four geodesic equations, therefore the equations are completely integrable.
As a consequence, the Hamiltonian principal function $S$ can be separated as
\be
S=\frac{1}{2}m^2\sigma-E t+L\phi+S_r(r)+S_\theta(\theta),
\ee
where $S_r$ and $S_\theta$ are the functions of $r$ and $\theta$, respectively. It follows that we can express $S_r(r)$ and $S_\theta(\theta)$ as
\be
S_r(r)=\int^r p_r dr=\int^r\frac{\sqrt{\mathcal{\hat{R}}}}{\Delta}dr,
\ee
\be
S_\theta(\theta)=\int^\theta p_\theta d\theta=\int^\theta \sqrt{\hat{\Theta}}\, d\theta,
\ee
where we have introduced
\bea
\hat{\mathcal{R}}&=&E^2 r^4-\Delta(m^2 r^2+Q+L^2),\quad \Delta=r^2-\frac{r_0^{D-3}}{r^{D-5}},\\
\hat{\Theta}&=&Q-\cot^2\theta \ L^2.
\eea
The functions $\hat{R}(r)$ and $\hat{\Theta}(\theta)$ are generally called the radial and angular ``potentials''.
Moreover, the conditions $\mathcal{\hat{R}}\ge0$ and $\hat{\Theta}\ge0$ must hold due to their appearances in the square roots. The roots of $\mathcal{\hat{R}}=0$ and $\hat{\Theta}=0$ give the turning points in the radial and angular directions, respectively.
Taking the variation of the Hamiltonian principal function $S$ with respect to $\kappa$, $E$ and $L$, separately,
we obtain \cite{Carter:1968rr}
\bea\label{er}
\int^r\frac{dr}{\sqrt{\mathcal{\hat{R}}}}&=&\int^\theta\frac{d\theta}{\sqrt{\hat{\Theta}}},\\\label{et}
t&=&\int^r\frac{Er^4}{\Delta\sqrt{\mathcal{\hat{R}}}}dr,\\\label{ephi}
\phi&=&\int^r\frac{L}{\sqrt{\mathcal{\hat{R}}}}dr+\int^\theta\frac{\cot^2\theta L}{\sqrt{\hat{\Theta}}}d\theta.
\eea
In fact, these integrals are the path integrals along the trajectory connecting the two points. Here we use the notation \lq\lq{}$\int$\rq\rq{} for simplicity. In general, these integrals cannot be evaluated analytically. In this paper, we consider the geodesic motion of null particle, i.e. $m=0$ and our task is to evaluate the radial integrals analytically starting from the near-horizon region and ending in the far region at the leading order in the large $D$ limit.

In the large $D$ limit, the near-horizon region is defined by $r-r_0\ll r_0$ and the metric takes the form
\be
ds^2=-\left(1-\frac{1}{\sR}\right)dt^2+\frac{r_0^2}{D^2}\frac{d\sR^2}{\sR(\sR-1)}
+r_0^2(d\theta^2+\sin^2\theta d\phi^2+\cos^2\theta d\Omega^2_{D-4}).
\ee
to the leading order in $1/D$. In the above, in order to  describe the near-horizon region appropriately  we have introduced
a new radial coordinate
\be\label{bd}
\sR=\left(\frac{r}{r_0}\right)^D,
\ee
in terms of  which the  near-horizon region  is given by $\log \sR\ll D$.

On the other hand, according to the property of the large $D$ limit,  the far region is defined to be  $r-r_0\gg r_0/D$ or  is given by $\log \sR\gg1$ in terms of $\sR$. Its metric describes just the flat spacetime
\be
ds^2=-dt^2+dr^2+r^2 (d\theta^2+\sin^2\theta d\phi^2+\cos^2\theta d\Omega^2_{D-4}).
\ee
The two regions overlap in $1\ll\log\sR\ll e^D$, which allows us to solve the geodesic equations by using the method of matched asymptotic expansions (MAE) for all geodesics extending from the near-horizon region to the far region.

From Eq. (\ref{cq}), we find that $Q\ge0$ is always true at every point of the
trajectory. Furthermore, $Q=0$ corresponds the trajectories on the equatorial plane where $\theta=\pi/2$. 
It is convenient to introduce two  rescaled  parameters
\bea\label{np}
q=\frac{\sqrt{Q}}{E},\quad\lambda=\frac{L}{E},
\eea
to describe the trajectory of the photon. For the Schwarzschild black holes, due to the spherical symmetry one can arbitrarily choose the position of the equator. Without loss of generality
one can always fix the trajectory of the photon onto the equator so that $\theta=0$ and $Q=0$. In this case, as we will see below the geodesic equations can be integrated as algebraic equations. From Eqs. (\ref{er}),  (\ref{et}) and (\ref{ephi}) we obtain
\bea\label{ner}
r&=&\int^r\frac{dr}{\sqrt{\mathcal{R}}},\label{net}\\
t&=&\int^r\frac{r^4}{\Delta\sqrt{\mathcal{R}}}dr,\\\label{nephi}
\phi&=&\int^r\frac{\lambda}{\sqrt{\mathcal{R}}}dr,
\eea
where
\bea
\mathcal{R}(r)&=&r^4-\Delta\lambda^2\label{sR-Sch}.
\eea
As mentioned above, we are interested in the null geodesics which originate from $(t_n, r_n, \theta_n, \phi_n)$ with $r_n-r_0\ll r_0$ in the near-horizon region and reach out to $(t_f, r_f, \theta_f, \phi_f)$ with $r_f-r_0\gg r_0/D$ in the far region. Generally, these integrals are of elliptic type so one cannot obtain them analytically. Fortunately, the method of MAE in the large $D$ limit will help us to evaluate the radial integrals tactfully. More specifically, we will perform the radial integrals in the near and far region separately, and match their integration constants in the overlap region.

\subsection{The radial motion}
In this subsection, we focus on the radial integral as follows
\bea
I=\int^{r_f}_{r_n}\frac{dr}{\sqrt{\mathcal{R}}},
\eea
to the leading order in $1/D$ via MAE. In the following, to simplify our expression we will set $r_0=1$ without loss of generality. From (\ref{sR-Sch}), we find that in the near-horizon region $\mathcal{R}$ can be rewritten in terms of $\sR$ as
\be
\mR_n=1-\left(1-\frac{1}{\sR}\right)\lambda^2,
\ee
while the far region is of the form
\bea
\mathcal{R}_f=r^4-r^2\lambda^2.
\eea
We are considering the case that the geodesics originate from the neighborhood of the event horizon and extend through to the infinity, that is to say, there is no turning point outside the event horizon. Recall that turning points in the radial motion is given by $\mR_f=0$. In the far region,
they are at
\bea
r_{f\pm}=\pm\lambda.
\eea
Obviously, one can hide the turning points behind the horizon when the condition \footnote{We are thankful to anonymous referee for pointing out that this condition is similar to the one under which the absorption rate of the massless scalar field  in the background of Schwarzschild black hole becomes almost 1, when the frequency is much
higher than the angular momentum, that is, the black hole is an almost perfect absorber in this case. The detailed discussion can be found in \cite{Emparan:2013moa}.    }
\be\label{tc1}
\lambda<1,
\ee
holds. In addition, in the near-horizon region, the turning point is at
\bea
\sR_n=\left(\frac{\lambda^2}{\lambda^2-1}\right).
\eea
Combining with the condition (\ref{tc1}), one can immediately find
\be
\sR_n<0,
\ee
which implies the turning point in the near-horizon region is always inside the event horizon. Thus, we conclude that the condition (\ref{tc1}) sufficiently enables that the geodesics sending from the horizon will not bounce back before they reach the far region. In the following we will consider only the photons which satisfy (\ref{tc1}).

First of all, in the near-horizon region, to the leading order in $1/D$ we find that the radial integral is given analytically by
\bea
I_n(\sR)=\frac{1}{D\eta}\log\left[\lambda^2+2\eta\sR(\eta+\sqrt{\mR_n(\sR)})\right]+C_n,
\eea
where we have introduced a new parameter $\eta=\sqrt{1-\lambda^2}$ to simplify the expression and $C_n$ is an integration constant. In the far region, the analytical result of the integral is given by
\bea
I_f(r)=-\frac{\arctan\left(\frac{\lambda}{r^2-\lambda^2}\right)}{\lambda}+C_f,
\eea
with  $C_f$ being an integration constant. The relation between $C_n$ and $C_f$ is determined by matching $I_n=I_f$ in the overlap region, i.e. $1\ll\log \sR\ll D$. To achieve this, we firstly expand $I_n$ at large $\sR$, the solution to the leading order in $1/\sR$ takes the form
\bea
I_n(\sR)=\frac{\log \sR+\log(4\eta^2)}{D\eta}+C_n.
\eea
For the integral in the far region, due to $\log \sR\ll D$, which allows us to expand $ r=1+\frac{1}{D}\log\sR+\mathcal{O}(D^{-2})$,
we find the asymptotic behavior of $I_f(r)$ in the overlap region
\be
I_f(\sR)=-\frac{\arctan(\lambda/\eta)}{\lambda}+\frac{\log\sR}{D\eta}+C_f.
\ee
Matching $I_n=I_f$ in the overlap region gives us
\be
C_f=\frac{\arctan(\lambda/\eta)}{\lambda}+\frac{\log(4\eta^2)}{D\eta}+C_n.
\ee
Then the complete radial integral of the photons extending from the near-horizon region to the far region $I=I_f(r_f)-I_n(\sR_n)$ is determined to be
\bea
I&=&-\frac{\arctan\left(\lambda/\sqrt{r_f^2-\lambda^2}\right)}{\lambda}-\frac{1}{D\eta}\log\left[\lambda^2+2\eta\sR_n(\eta+\sqrt{\mR_n(\sR_n)})\right]\\
&&+\frac{\arctan(\lambda/\eta)}{\lambda}+\frac{\log(4\eta^2)}{D\eta}.\nn
\eea
The expression could be simplified as
\bea\label{sr}
I=-\frac{1}{\lambda}\arctan\frac{\lambda\eta-\lambda D_f}{\lambda^2+\eta D_f}-\frac{1}{D\eta}\log\left(\frac{\lambda^2+2\eta \sR_n(\eta+D_n)}{4\eta^2}\right),
\eea
where we have defined
\bea
D_n=\sqrt{\mR_n(\sR_n)},\quad D_f=\frac{1}{r_f}\sqrt{\mR_f(r_f)}.
\eea

\subsection{The $r$-$\phi$ and $r$-$t$ motion}
In this subsection, we show the analytical results of the radial integrals for the $r$-$\phi$ and $r$-$t$ motion, which are denoted by
\bea\label{sphi}
I^\phi&=&\lambda\int^{r_f}_{r_n}\frac{dr}{\sqrt{\mR(r)}}=\lambda I,\\
I^t&=&\int^{r_f}_{r_n}\frac{r^4dr}{\Delta\sqrt{\mR(r)}}.
\eea
The result of $I$ in the previous subsection applies to $I^\phi$, which leaves the integral $I^t$ to be evaluated. Through similar computation, we find that in the near-horizon region
\bea
I^t_n(\sR)=\frac{1}{D}\left(\log\frac{\sR-1}{\lambda^2+\sR\left(1+\eta^2+2\sqrt{\mR_n(\sR)}\right)}
+\frac{\lambda^2+2\eta\sR\left(\eta+\sqrt{\mR_n(\sR)}\right)}{\eta}\right)+C_n^t,
\eea
 and   in the far region
\bea
I^t_f(r)=\sqrt{r^2-\lambda^2}+C_f^t,
\eea
where $C_{n,f}^t$ are integration constants. In the overlap region, their asymptotic behaviors are respectively
\bea
I^t_n(\sR)&=&\frac{\log \sR+\log(4\eta^2)}{D\eta}-\frac{2\log(1+\eta)}{D}+C_n^t,\\
I^t_f(\sR)&=&\eta+\frac{\log\sR}{D\eta}+C_f^t.
\eea
Then matching $I_n^t=I_f^t$ in the overlap region we obtain
\bea
C_f^t=\frac{\log(4\eta^2)}{D\eta}-\frac{2\log(1+\eta)}{D}-\eta+C_n^t,
\eea
thus, from $I^t=I_f^t(r_f)-I^t_n(\sR_n)$, we find
\bea\label{st}
I^t=\sqrt{r^2-\lambda^2}-\eta-\frac{1}{D}\log\frac{(\sR_n-1)(1+\eta)^2}{\lambda^2+\sR_n(1+\eta^2+2D_n)}
-\frac{1}{D\eta}\log\frac{\lambda^2+2\eta\sR_n(\eta+D_n)}{4\eta^2},
\eea
where $D_n$ is the same as the one in Eq. (\ref{sr}).

\section{Singly-spinning Myers-Perry Black Hole}\label{MP_BH}

In this section, we proceed to perform the radial integrations of the null geodesics in the background of a singly-spinning Myers-Perry(MP) black hole, which is more attractive. When the rotation parameter $a$ goes to zero, it will reduce to the case of the Schwarzschild black hole in the previous section.

The metric of a $D$ dimensional asymptotically flat MP black hole spinning in a single plane is given by \cite{Myers:1986un}
\be
ds^2=-\frac{\Delta}{\Sigma}(dt-a\sin^2 \theta d\phi)^2+\frac{\sin^2\theta}{\Sigma}\left[(r^2+a^2)d\phi-adt\right]^2
+\frac{\Sigma}{\Delta}dr^2+\Sigma d\theta^2
+r^2\cos^2\theta d\Omega^2_{D-4},
\ee
where
\be
\Sigma=r^2+a^2 \cos^2\theta,\quad
\Delta=r^2+a^2-\frac{r_0^{D-3}}{r^{D-5}}.
\ee
This solution of the Einstein vacuum equations is characterized by two parameters,  the
mass-radius $r_0$ and the rotation parameter $a$,
\be
M=\frac{(D-2)r_0^{D-3}\Omega_{D-2}}{16\pi G},\quad
a=\frac{D-2}{2}\frac{J}{M},
\ee
where $\Omega_{D-2}=2\pi^{(D-1)/2}/\Gamma[(D-1)/2]$ is the volume of a unit $(D-2)$ sphere,
$M$ and $J$ are the ADM mass and angular momentum, respectively. The
 event horizon corresponds to the largest real root $r=r_+$ of $\Delta(r)=0$, i.e.
\be
r_+^2+a^2-\frac{r_0^{D-3}}{r_+^{D-5}}=0.\label{Delta_MP}
\ee
For $D\geq6$, $\Delta(r)$ always has a positive real root for all values of $a$ and therefore exists
the black hole with arbitrary large angular momentum per unit mass. From Eq. (\ref{Delta_MP}) we can see that the horizon radius is given by
\be
r_+=r_0\left(1+\frac{a^2}{r_+^2}\right)^{-\frac{1}{D-3}},
\ee
thus when $D\to\infty$, it follows that
\be
r_+\to 1,\quad r_+^D\to \frac{1}{1+a^2},
\ee
where we have set $r_0=1$.
From above, we may still define the near-horizon region as $r-r_0\ll r_0$ and the far region as $r-r_0\gg \frac{r_0}{D}$, or in
terms of  the radial coordinate $\sR$, they are given by $\log \sR\ll D$ and $\log \sR\gg 1$. In addition, the horizon is located at $\sR_+=1/(1+a^2)$ using the coordinate $\sR$.

Similar to the case of the Schwarzschild black hole, from the Hamilton-Jacobi formalism, the null geodesic equations can be integrated as
\footnote{The general discussions on the integrability of null geodesics in higher dimensional rotating black hole spacetime can be found in \cite{Vasudevan:2004mr, Page:2006ka,Frolov:2017kze, Lunin:2017drx}.  }
\bea\label{krr}
0&=&\int^r\frac{dr}{\sqrt{\mR}}-\int^\theta\frac{d\theta}{\sqrt{\Theta}},\\\label{ktt}
t&=&\int^\theta\frac{a^2\cos^2\theta}{\sqrt{\Theta}}d\theta
+\int^r\frac{r^2(a^2+r^2)+a(\lambda-a)(\Delta-r^2-a^2)}{\Delta\sqrt{\mR}}dr,\\\label{kphi}
\phi&=&\int^r\frac{ar^2+(\lambda-a)(\Delta-a^2)}{\Delta\sqrt{\mR}}dr
+\int^\theta\frac{\cot^2\theta \lambda}{\sqrt{\Theta}}d\theta,
\eea
where
\bea\label{kr}
\mR&=&[( r^2+a^2)-a\lambda]^2-\Delta(q^2+(\lambda-a)^2),\\
\Theta&=&q^2-\cos^2\theta(\lambda^2\csc^2\theta-a^2),
\eea
where we  have  also  used the rescaled quantities $(q, \lambda)$ defined in Eq. (\ref{np}), since the geodesic of the photon is independent of its energy. Similarly, the functions $\hat{R}(r)$ and $\hat{\Theta}(\theta)$ represent the radial and angular potentials. Their zeros correspond to the turning points in the trajectories. To guarantee that the geodesic can be extended to the infinity from the near-horizon region, the condition $\mR>0$ outside the event horizon must be respected. In the following we will check this point in the near-horizon and far regions, respectively.

\subsection{$r$-$\theta$ motion}
\label{secr}
In this subsection, we will perform the radial integration using the method of MAE carefully.  Let us  define
\bea
I=\int^{r_f}_{r_n}\frac{dr}{\sqrt{\mathcal{R}}},
\eea
where for the rotating black holes, $\mR$ is given by Eq. (\ref{kr}). In the near-horizon region, taking $\sR$ as the variable, up to the leading order in $1/D$, we have
\be
\mR_n=[(1+a^2)-a\lambda]^2-\Delta(q^2+(\lambda-a)^2)],\quad \Delta=1+a^2-\frac{1}{\sR}.
\ee
Recall that the condition $\mR_n>0$ should be always satisfied outside the event horizon, which leads to
\be\label{SMPbound}
q^2<1-\frac{\lambda^2}{a^2+1}.
\ee
That is to say, the photons under this constraint originating from the outer horizon will not bounce back in the outward radial motion.

In the far region, the metric is simplified to be
\be
ds^2=-dt^2
+\frac{r^2+a^2 \cos^2\theta}{r^2+a^2}dr^2+(r^2+a^2 \cos^2\theta) d\theta^2
+(r^2+a^2)\sin^2\theta d\phi^2
+r^2\cos^2\theta d\Omega^2_{D-4}.
\ee
This is just the flat spacetime in the ellipsoidal coordinates, which brings difficulty to the integrations of the geodesic equations as we will see below. In the far region the radial potential becomes
\bea\label{farR_SMP}
\mR_f=r^4+r^2 \left(a^2-\lambda ^2-q^2\right)-a^2 q^2,
\eea
then the integration in this region is given by
\bea
I_f=\int^{r_f}\frac{dr}{\sqrt{\mR_f}}.
\eea
Unfortunately, this integral in general is of elliptic type so cannot be obtained analytically.\footnote{We find that even though the integration could be  expressed in terms of the Elliptic integral, with which we cannot do the matching analytically in the overlapping region.} The analytically solvable case occurs when we take $a=0$ or $q^2=0$, corresponding to the case that the black hole is static or the geodesics are confined to the equatorial plane of the rotational black hole respectively. On the other hand, according to the linear analysis of the gravitational perturbation of the Myers-Perry black hole in the large $D$ limit \cite{Suzuki:2015iha}, the black hole becomes dynamically unstable when $a>1$, thus
we consider the case $a\leq 1$. In this case, due to the constraint (\ref{SMPbound}), we know that both $|a^2-q^2-\lambda^2|$ and $a^2q^2$ are smaller than 1.
Therefore, in the integral, for $r-1\ll 1$, $\mR_f$ can be well approximated by
\bea\label{farRappr_SMP}
\mR_f&\simeq& r^4+\left(a^2-\lambda ^2-q^2\right)r^2 -a^2 q^2 r^2\nn\\
&=&r^2(r^2-\xi_3),
\eea
where we have introduced
\be\label{xi3}
\xi_3=\lambda^2+q^2-a^2+a^2q^2.
\ee
Note that for $r-1\gg 1$, the first term in (\ref{farR_SMP}) would be dominant, then the above approximation still works well.
For the intermediate region of the integral domain $r-1\sim \mc O(1)$, in (\ref{farR_SMP}) the first term is larger than the other two terms, so it might be fine to use the above approximation. In fact, the numerical computation shows that the approximation (\ref{farRappr_SMP}) reproduces the result up to percent level accuracy.

Furthermore, recall that the turning points are given by the roots of $\mR_f=0$, and we let the maximum root be less than the horizon radius. Interestingly, when $\xi_3>0$, we find the same condition as  (\ref{SMPbound}) found in the near-horizon region. Actually, Eq.  (\ref{SMPbound}) leads to
\be
0<\xi_3<1.
\ee
 Thus, we conclude that when the condition (\ref{SMPbound}) holds, the null particle leaving from the near region can escape to the far region without hindrance.
In contrast, there is no turning points when $\xi_3<0$, and $\mR_f>0$ always holds. In this case, we find a new constraint on $(q,\lambda)$
\bea
q^2<\frac{a^2-\lambda^2}{1+a^2}.
\eea
Note that this is more stringent than the one given in the near-horizon region (\ref{SMPbound}).
In addition, we need to emphasize that if the geodesics are on the equatorial plane of a rotating black hole, our approximation is able to give exact results, as the term $a^2q^2$ is dropped in this special case.

We can perform the integration in the near-horizon region analytically
\bea
I_n&=&\int^{r_n}\frac{dr}{\sqrt{\mR_n}}\nonumber\\
&=&\frac{1}{D}\frac{\log \left(2 \sR_n \xi_1
\sqrt{\mR_n(\sR_n)}
+2\sR_n \xi_1^2+\xi_2^2\right)}{E\xi_1}+C_n,
\eea
where we have introduced
\be
\xi_1=\sqrt{\left(a^2+1\right) \left(1-q^2\right)-\lambda ^2},
\ee
\be
\xi_2=\sqrt{(a-\lambda )^2+q^2}.
\ee
Note that $\xi_1>0$ can be guaranteed by the requirement of  no turning point outside the event horizon for the photon, and it is related to $\xi_3$ by the identity
\be
\xi_1^2+\xi_3=1.
\ee
In the overlap region $1\ll\log \sR\ll D$, $I_n$ becomes
\be
I_n=\frac{1}{D}\frac{\log \sR }{
\xi_1}
+\frac{1}{D}\frac{2\log \left(2  \xi_1\right)}{
\xi_1}+C_n,
\ee
to the leading order in $1/\sR$.

On the other side, for the integral in the far region, the integral can be analytically obtained as well under the approximation \eqref{farRappr_SMP}.  Depending on the sign of the value $\xi_3$ takes, the integral takes different form.  If $\xi_3<0$,
the integral is obtained as
\bea
I_f^{(1)}(r)=\frac{1}{\xi_4}\log\frac{r}{\xi_4(\xi_4+\sqrt{r^2+\xi_4^2})}+C_f^{(1)},
\eea
where we have introduced a new parameter
\bea\label{xi4}
\xi_4\equiv\sqrt{|\xi_3|},
\eea
which is related to $\xi_1$ by $\xi_1^2-1=\xi_4^2$.

The other case is $0<\xi_3<1$, in this case the constraint on $(q,\lambda)$ becomes
\be
\frac{a^2-\lambda^2}{1+a^2}<q^2<1-\frac{\lambda^2}{1+a^2},
 \ee
 and the identify $1-\xi_1^2=\xi_4^2$ follows immediately. In this case, the analytical result of the radial integral in the far region takes the form
\be
I_f^{(2)}=-\frac{1}{\xi_4}\arctan\frac{\xi_4}{\sqrt{r^2-\xi_4^2}}+C_f^{(2)}.
\ee
Thus, following the same procedure where we have shown in the previous section, in the overlap region, $I_f$ becomes
\be
I_f^{(1)}=\frac{1}{ D } \frac{\log \sR}{\xi_1}-\frac{\log (\xi_4^2+\xi_1\xi_4)}{\xi_4}+C_f^{(1)},
\ee
and
\be
I_f^{(2)}=\frac{1}{D}\frac{\log \sR}{\xi_1}-\frac{\arctan\frac{\xi_4}{\xi_1}}{\xi_4}+C_f^{(2)}.
\ee
Matching $I_n = I_f$ in the overlap region we find
\be
C_f^{(1)}=C_n+\frac{\log (\xi_4^2+\xi_1\xi_4)}{\xi_4}+\frac{1}{D}\frac{\log \left(4  \xi_1^2\right)}{
\xi_1},
\ee
and
\be
C_f^{(2)}=C_n+\frac{\arctan\frac{\xi_4}{\xi_1}}{\xi_4}+\frac{1}{D}\frac{\log \left(4  \xi_1^2\right)}{
\xi_1}.
\ee
Then the integration $I=I_f(r_f)-I_n(\sR_n)$ is given by
\bea\label{Ir1}
I^{(1)}=\frac{1}{\xi_4}\log\frac{r_f(\xi_1+\xi_4)}{\xi_4+\sqrt{r_f^2+\xi_4^2}}
+\frac{1}{D\xi_1}\log\frac{4\xi_1^2}{2\sR_n\xi_1\sqrt{\mR_n(\sR_n)}+2\sR_n\xi_1^2+\xi_2},
\eea
and
\bea\label{Ir2}
I^{(2)}=\frac{1}{\xi_4}\arctan\frac{\xi_4\sqrt{r_f^2-\xi_4^2}-\xi_1\xi_4}{\xi_1\sqrt{r_f^2-\xi_4^2}
+\xi_4^2}+\frac{1}{D\xi_1}\log\frac{4\xi_1}{2\sR_n\xi_1\sqrt{\mR_n(\sR_n)}+2\sR_n\xi_1^2+\xi_2^2}.
\eea

\subsection{The $r$-$\phi$ and $r$-$t$ motion}

In this subsection, we turn our attention to the radial integrals of the $r$-$\phi$ and $r$-$t$ motion, under the approximation \eqref{farRappr_SMP}. First, we focus on the radial integral of the $r$-$\phi$ motion which can be rewritten as
\bea\label{Iphi}
I^\phi&=&\int^{r}\frac{ar^2+(\lambda-a)(\Delta-a^2)}{\Delta\sqrt{\mR(r)}}dr\nn\\
&=&\int^r dr\left[\frac{ar^2-(\lambda-a)a^2}{\Delta\sqrt{\mR(r)}}+\frac{\lambda-a}{\sqrt{\mR(r)}}\right].
\eea
Since $(\lambda-a)$ appearing in the second term is a constant, the result of this part is straightforward given by $(\lambda-a) I$, so only the first term remains to be integrated. In fact, we find it not hard to evaluate the integral of the first term in the near-horizon region and the result of $I_n^{\phi}$ takes
\bea\label{Iphii}
I_n^\phi&=&\frac{a}{D(1+a^2)}\log\frac{(1+a^2)\sR-1}{2\sR\sqrt{\mR_n}(1-a\lambda+a^2)+\xi_2^2+2\sR\xi_1^2+(1+a^2)\sR\xi_2^2}\nn\\
&&+\frac{\lambda}{D(1+a^2)\xi_1}\log\left(2\sR\xi_1^2+\xi_2^2+2\sR\xi_1\sqrt{\mR_n}\right)+C_n^\phi.
\eea
In the overlap region, $I_n^\phi$ becomes
\be
I_n^\phi=\frac{1}{D}\frac{\lambda }{(a^2+1)\xi_1}\log \sR+\frac{1}{D}\frac{1}{(a^2+1)\xi_1}
\left(2\lambda\log(2\xi_1)+a\xi_1\log\frac{1+a^2}{(1+a^2-a\lambda+\xi_1)^2}\right)+C_n^\phi.
\ee
In the far region, however, the integration of the first term in Eq. (\ref{Iphi}) cannot be computed directly. To keep the paper more logical, we leave the computational details to the appendix \ref{appA}, and present the results as below
\bea\label{I1}
I^{\phi(1)}_f=\frac{\lambda}{\sqrt{a^2-\xi_4^2}}\tan ^{-1}\left(\frac{\sqrt{\xi_4^2+r^2}}{\sqrt{a^2-\xi_4^2}}\right)
+C_f^{\phi(1)},
\eea
for $q^2< \frac{a^2-\lambda^2}{1+a^2}$, and
\be\label{I2}
I^{\phi(2)}_f=\frac{\lambda  }{\sqrt{a^2+\xi_4^2}}\tan ^{-1}\left(\frac{\sqrt{r^2-\xi_4^2}}{\sqrt{a^2+\xi_4^2}}\right)+C_f^{\phi(2)},
\ee
for $\frac{a^2-\lambda^2}{1+a^2}<q^2<1-\frac{\lambda^2}{1+a^2}$ . In the overlap region, $I_f^\phi$ becomes
\be
I^{\phi(1,2)}_f=\frac{\lambda  }{D\left(a^2+1\right)  \xi_1}\log \sR+\frac{\lambda  }{\sqrt{a^2-\xi _1^2+1}}\tan ^{-1}\left(\frac{\xi_1}{\sqrt{a^2-\xi_1^2+1}}\right)+C_f^{\phi(1,2)},
\ee
from which we see that the difference between $I^{\phi(1)}$ and $I^{\phi(2)}$ only comes from the integration constant. Matching $I_n^\phi=I_f^\phi$ in the overlap region we find
\bea
C_f^{\phi(1,2)}&=&C^\phi_n+\frac{1}{D}\frac{1}{(a^2+1)\xi_1}
\left(2\lambda\log(2\xi_1)+a\xi_1\log\frac{1+a^2}{(1+a^2-a\lambda+\xi_1)^2}\right)\nonumber\\
&&-\frac{\lambda  }{\sqrt{a^2+1-\xi_1^2}}\tan ^{-1}\left(\frac{ \xi_1}{\sqrt{a^2+1-\xi_1^2}}\right).
\eea
To summarize, the integral $I^\phi=I^{\phi}_f(r_f)-I^\phi_n(\sR_n)$ is given by
\bea\label{Iphi1}
I^{\phi(1)}&=&\frac{\lambda}{\sqrt{a^2-\xi_4^2}}\tan^{-1}\frac{\sqrt{a^2-\xi_4^2}\left(\sqrt{r_f^2+\xi_4^2}-\xi_1\right)}{a^2-\xi_4^2+\xi_1\sqrt{r_f^2+\xi_4^2}}\nn\\
&&+\frac{a}{D(a^2+1)}\log\frac{\left(1+a^2\right)\left[2\sR_n\sqrt{\mR_n}(1-a\lambda+a^2)+\xi_2^2+2\sR_n\xi_1^2+(1+a^2)\sR_n\xi_2^2\right]}{(1+a^2-a\lambda+\xi_1)^2\left[(1+a^2)\sR_n-1\right]}\nn\\
&&+\frac{\lambda}{D(a^2+1)\xi_1}\log\frac{4\xi_1^2}{2\sR_n\xi_1^2+\xi_2^2+2\sR_n\xi_1\sqrt{\mR_n}},
\eea
and
\bea\label{Iphi2}
I^{\phi(2)}&=&\frac{\lambda}{\sqrt{a^2+\xi_4^2}}\tan^{-1}\frac{\sqrt{a^2+\xi_4^2}\left(\sqrt{r_f^2-\xi_4^2}-\xi_1\right)}{a^2+\xi_4^2+\xi_1\sqrt{r_f^2-\xi_4^2}}\nn\\
&&+\frac{a}{D(a^2+1)}\log\frac{\left(1+a^2\right)\left[2\sR_n\sqrt{\mR_n}(1-a\lambda+a^2)+\xi_2^2+2\sR_n\xi_1^2+(1+a^2)\sR_n\xi_2^2\right]}{(1+a^2-a\lambda+\xi_1)^2\left[(1+a^2)\sR_n-1\right]}\nn\\
&&+\frac{\lambda}{D(a^2+1)\xi_1}\log\frac{4\xi_1^2}{2\sR_n\xi_1^2+\xi_2^2+2\sR_n\xi_1\sqrt{\mR_n}}.
\eea

Similarly, the radial integral in the $r$-$t$ motion is given by
\bea\label{rt_MP}
I^t = \int^{r_f}_{r_n} \frac{ r^2 (a^2 + r^2) + a (\lambda - a ) (\Delta - r^2
   - a^2)}{\Delta \sqrt{\mR}} d r.
\eea
In the near-horizon region, in terms of the new radial coordinate $\sR$, the
integral becomes
\begin{eqnarray*}
  I^t_n & = & \frac{1}{D} \log \frac{(a^2 + 1)\sR - 1}{\sR (a^2 - a \lambda + 1)
  \left( a^2 - a \lambda + 1 + 2 \sqrt{\mR_n(R)} \right)+\sR \xi_1^2 +
  \xi_2^2}\\
  &  & + \frac{1}{D} \frac{\log \left(2\sR \xi_1 \sqrt{\mR_n (\sR)} + \sR
  \xi_1^2 + \xi_2^2\right)}{\xi_1}+C^t_n .
\end{eqnarray*}
In the overlap region, it becomes
\bea
I_n^t = \frac{1}{D \xi_1} \log R + \frac{1}{D \xi_1} \left( 2 \log (2
   \xi_1) + \xi_1 \log \frac{1 + a^2}{(a^2 - a \lambda + 1)^2} \right) .
\eea
In the far region, the integral (\ref{rt_MP}) becomes
\bea
 I^t_f = \sqrt{r^2 + (a^2
   - q^2 - \lambda^2 - a^2 q^2)} + C^t_f = \sqrt{r^2 + \xi_1^2} + C^t_f .
\eea
In the overlap region, it becomes
\be
I^t_f = \frac{1}{D \xi_1} \log R + \xi_1 + C^t_f .
\ee
Matching $I_n^t = I^t_f$ in the overlap region, we have
\be
C^t_f = C^t_n + \frac{1}{D \xi_1} \left( 2 \log (2 \xi_1) + \xi_1 \log
   \frac{1 + a^2}{(a^2 - a \lambda + 1)^2} \right) - \xi_1 .
\ee
To summarize, the integral of the $r$-$t$ motion is given by
\bea\label{Ikt}
I^t&=&\sqrt{r_f^2+\xi_1^2}-\xi_1+\frac{1}{D\xi_1}\log\frac{4\xi_1^2}{2\sR \xi_1 \sqrt{\mR_n (\sR_n)} + \sR_n\xi_1^2 + \xi_2^2}\\
&&+\frac{1}{D}\log\frac{\left(1+a^2\right)\left[\sR_n (a^2 - a \lambda + 1)
  \left( a^2 - a \lambda + 1 + 2 \sqrt{\mR_n(R)} \right)+\sR_n \xi_1^2 +
  \xi_2^2\right]}{\left(a^2 - a \lambda + 1)^2\right)\left[(a^2 + 1)\sR_n - 1\right]}.\nn
\eea
Hereto, we have found all the radial integrals in the null geodesic Eqs. (\ref{krr}), (\ref{ktt}), (\ref{kphi}) analytically.

It is remarkable that even though the above analytical expressions are obtained under the approximation (\ref{farRappr_SMP}), the expressions are exact when we restrict the geodesics onto the equatorial plane, i.e. $\theta=\pi/2$. In this case, the $\theta$ integrals can be dropped \footnote{In this case, we also find that Eq. (\ref{krr}) is ineffective since the conserved quantity $\kappa\equiv L^2$ is no longer an independent constant and the Carter constant $Q=0$ (or $q^2=0$).}.  Therefore, substituting $q^2=0$ in Eqs. (\ref{Ir1}), (\ref{Ir2}), (\ref{Iphi1}), (\ref{Iphi2}) and (\ref{Ikt}), the resulting algebraic expressions can simply give us the the time and azimuthal angle along the null geodesic from the starting point $(t_n, \sR_n, \phi_n)$ in the near-horizon region to the endpoint $(t_f, r_f, \phi_f)$ in the far region outside the black hole, respectively. The results are straightforward and lengthy so we omit the concrete expressions here to keep the writing dapper.

\section{Summary}\label{summary}

In this paper, we studied the null geodesics for the massless particles in the background of the Schwarzschild and the singly-spinning Myers-Perry black holes in the large dimension limit. In this particular limit, by using the method of matched asymptotic expansions (MAE), we were able to integrate analytically all the radial integrals in Carter\rq{}s integral geodesic equations to the leading order in $1/D$.

We first considered the Schwarzschild black hole, which can be taken as the special  $a=0$ limit of the Myers-Perry black holes. Due to  its spherical symmetry, the radial integral could be  derived easily and the results  are Eqs. (\ref{sr}), (\ref{sphi}) and (\ref{st}).   Then, we generalized the study to the singly-spinning Myers-Perry black hole, where the rotation parameter $a$ is not fixed and limited by the upper bound $a\leq r_0$. Under the approximation \eqref{farRappr_SMP}, we found the analytical expressions of all radial integrals,  Eqs. (\ref{Ir1}), (\ref{Ir2}), (\ref{Iphi1}), (\ref{Iphi2}) and (\ref{Ikt}). One remarkable feature is that $I^r$ and $I^\phi$ take two different expressions depending on the impact parameters $q$ and $\lambda$ as well as the rotation parameter $a$. In contrast, they only differ by a constant factor in the  spherically symmetric  case. If the geodesics are confined to the equatorial plane connecting the start point $(t_n, \sR_n, \phi_n)$ near the horizon with the end point $(t_f, r_f, \phi_f)$ far from the black hole, the expressions will be simplified with $\theta=\pi/2, q=0$ and constant $\lambda$ along the trajectory.

The radial integrals obtained in the large dimension limit using the method of MAE allow us to investigate the influences of $a$ on the geodesics analytically in contrast to the previous work \cite{Porfyriadis:2016gwb}, where the radial integrals were obtained only in the extremal Kerr case.  Despite the fact that our present formulas only have sufficient accuracy when the dimension becomes large enough, one may find some universal behaviours under the influence of the rotation parameter $a$.



Let us conclude this paper with some outlooks. First of all, our proposal can be extended to the modified gravity easily as long as the four conserved constants of the geodesic motion exist, no matter the black hole is static or rotational, as well as neutral or charged. One may find some more interesting features beyond the Einstein gravity and put forward an effective model to characterize the black holes in reality. Secondly, one can apply our result to investigate the observational quantities of hot spots or electromagnetic radiation analytically, and see if there exists universal behavior independent of the spacetime dimension. Finally it would be interesting to push our study to high orders in $1/D$ so that one can develop a perturbative framework in $1/D$ to study  four-dimensional black holes.


\section*{Acknowledgments}
The work was in part supported by NSFC Grant No. 11335012, No. 11325522, No. 11675015, No. 11735001, No. 11775022 and No. 11847241.

\appendix

\section{Computational detail of Eqs. (\ref{I1}) and (\ref{I2})}\label{appA}
In this appendix, we want to show some important computational details when deriving Eq. (\ref{I1}) and (\ref{I2}). In the far region, substituting $\Delta=r^2+a^2$ in Eq. (\ref{Iphi}), we obtain
\bea
I_f^\phi&=&\int^rdr\left(\frac{ar^2-(\lambda-a)a^2}{(r^2+a^2)\sqrt{\mR(r)}}+\frac{\lambda-a}{\sqrt{\mR(r)}}\right)\nn\\
&=&\int^rdr\frac{\lambda}{\sqrt{\mR(r)}}-\int^r\frac{\lambda a^2}{r(r^2+a^2)\sqrt{r^2-\xi_3}}.
\eea
Since we have worked out the first integral in Section \ref{secr}, the remaining part is given by
\bea\label{rxx}
I=\int^rdr\frac{1}{r(r^2+a^2)\sqrt{r^2-\xi_3}},
\eea
where, $\xi_3$ is given by (\ref{xi3}) and its sign is undetermined. We introduce a new variable $x=\sqrt{r^2-\xi_3}$, and Eq. (\ref{rxx}) can be rewritten as
\be
I=\int^{\sqrt{r^2-\xi_3}}\frac{dx}{(x^2+\xi_3+a^2)(x^2+\xi_3)},
\ee
where we have used $dr=\frac{xdx}{\sqrt{x^2+\xi_3}}$. Then, the integral can be performed easily: $\xi_3<0$,
\bea
I=\frac{\tan^{-1}\frac{\sqrt{r^2+\xi_4^2}}{a^2-\xi_4^2}}{a^2\sqrt{a^2-\xi_4^2}}+\frac{\tan^{-1}\frac{\sqrt{r^2+\xi_4^2}}{\xi_4}}{a^2\xi_4},
\eea
in which $\xi_4^2$ is given by (\ref{xi4}),  while for $\xi_3>0$,
\bea
I=\frac{\tan^{-1}\frac{\sqrt{r^2-\xi_4^2}}{a^2+\xi_4^2}}{a^2\sqrt{a^2+\xi_4^2}}+\frac{\tan^{-1}\frac{\sqrt{r^2-\xi_4^2}}{\xi_4}}{a^2\xi_4}.
\eea
Using these identities and after some tedious calculations, Eqs. (\ref{I1}) and (\ref{I2}) can be obtained.

\end{document}